\documentclass[preprint]{JHEP3}
\usepackage{amsmath}
\newcommand{\ep}{\varepsilon}
\newcommand{\Li}[2]{{\mbox{Li}}_{#1}\left(#2\right)}
\newcommand{\Cl}[2]{{\mbox{Cl}}_{#1}\left(#2\right)}

\newcommand{\LS}[3]{{\mbox{Ls}}_{#1}^{(#2)}\left(#3\right)}  
\newcommand{\LsLsc}[4]{{\mbox{LsLsc}}_{#1,#2,#3} \left(#4\right)}
\title{
Multiple (inverse) binomial sums of arbitrary weight and depth
and the all-order $\ep$-expansion of generalized hypergeometric functions
with one half-integer value of parameter
}
\author{
M.~Yu.~Kalmykov\thanks{Present Address: 
II. Institut f\"ur Theoretische Physik, Universit\"at Hamburg ,
Luruper Chaussee 149, 22761 Hamburg, Germany.
Email: \protect\email{kalmykov@theor.jinr.ru}}\\
Department of Physics, Baylor University, \\
One Bear Place, Box 97316 \\
Waco, TX 76798-7316 \\
\\
Bogoliubov Laboratory of Theoretical Physics, \\
Joint Institute for Nuclear Research, \\
$141980$ Dubna (Moscow Region), Russia \\
}
\author{
B.F.L.~Ward, \quad S.A.~Yost\thanks{Present address:
Department of Physics, Princeton University, Princeton, NJ 08540.
Email: \protect\email{syost@princeton.edu}} \\
Department of Physics, Baylor University, \\
One Bear Place, Box 97316 \\
Waco, TX 76798-7316}
\preprint{ 
 BU-HEPP-07-01}
\date{July 2007}
\abstract{ \\
We continue the study of the construction of analytical coefficients of the
$\ep$-expansion of hypergeometric functions and their connection with 
Feynman diagrams.  In this paper, we show the following results: \\

Theorem $A$: \\
The multiple (inverse) binomial sums 
\[
\sum_{j=1}^\infty \frac{1}{\binom{2j}{j}^k}\frac{z^j}{j^c} S_{a_1}(j-1) 
\cdots S_{a_p}(j-1) \; ,
\]
where $k = \pm 1$, $S_a(j)$ is a harmonic series, 
$S_a(j) = \sum_{k=1}^j \frac{1}{k^a}$, and 
$c$ is any integer number are expressible in terms of Remiddi-Vermaseren 
functions; \\

Theorem $B$: \\
The hypergeometric functions
\[
{}_pF_{p-1}(\vec{A} \!+\! \vec{a}\ep; 
             \vec{B} \!+\! \vec{b} \ep, \tfrac{1}{2} \!+\! B_{p-1}; z) \;,
\qquad
{}_pF_{p-1}(\vec{A} \!+\! \vec{a}\ep, \tfrac{1}{2} \!+\! A_{p} ; 
             \vec{B} \!+\! \vec{b} \ep ;z) \;,
\]
are expressible in terms of the harmonic 
polylogarithms of Remiddi and Vermaseren with coefficients that are 
ratios of polynomials. \\

}
\begin{document}
\section{Introduction}
Feynman diagrams\,\cite{Bogolyubov} are a primary tool for calculating 
radiative corrections to any processes within the Standard Model or its 
extensions.  With increasing of accuracy of measurements, more and more 
complicated diagrams (with increasing number of loops and legs and increasing 
numbers of variables associated with different particle masses) must
be evaluated. The essential progress in such calculations is often associated 
with the invention of new (mainly mathematical) algorithms 
({\it e.g.}  Refs.\ \cite{Smirnov,Tausk}) and their realization as a 
computer programs ({\it e.g.} Refs.\ \cite{Laporta,Anastasiou}). 
One fruitful approach to the calculation of Feynman diagrams is 
based on their representation in terms of hypergeometric 
functions\,\cite{BD} or multiple series\,\cite{FKV99,KV00}. 
We will refer to such representations as {\it hypergeometric representations}
for Feynman diagrams.  Unfortunately, there does not exist a universal 
hypergeometric representation for all types of diagrams.  Constructing these
representations is still a matter of the personal experience of the 
researcher \cite{hyper,DK01,JKV03,DK04,MKL06}. Nevertheless, existing 
experience with Feynman diagrams leads us to expect that all Feynman diagrams
should be associated with hypergeometric functions.  

For practical applications,
finding a hypergeometric representation is not enough. It is necessary to 
construct the so-called $\ep$-expansion, which we may understand as the 
construction of the analytical coefficients of the Laurent expansion of 
hypergeometric functions around rational values of their parameters. 
In this direction, very limited results are available.\footnote{
One of the {\it classical tasks} in mathematics is to find the full set of 
parameters and arguments for which hypergeometric functions
are expressible in terms of algebraic functions. Quantum field theory makes a 
{\it quantum} generalisation of this classical task: to find the full set of 
parameters and arguments so that the all-order $\ep$-expansion is 
expressible in terms of known special functions or identify the full set of 
functions which must be invented in order to construct the
all-order $\ep$-expansion of generalized hypergeometric functions.}
The pioneering systematic activity in studying the Laurent series expansion 
of hypergeometric functions at particular values of the argument 
$(z=1)$ was started by David Broadhurst\,\cite{Broadhurst:1996}
in the context of Euler-Zagier sums (or multidimensional 
zeta values) \cite{Euler-Zagier}. This activity has received further consideration for 
another, physically interesting point, $z=1/4$ (see the relevant Appendix in 
Ref.\ \cite{KV00,DK01}), and also for the ``primitive sixth roots of unity''
(see Ref. \cite{Broadhurst:1998}). 
Over time, other types of sums\footnote{See Eq.~(\ref{binsum}) 
for clarifying of terminology.} have been analysed in a several publications:
{\it harmonic} sums\,\cite{FKV99,nested1}, 
{\it generalized harmonic sums}\,\cite{nested2,DK04},
{\it binomial} sums\,\cite{JKV03,DK04} and 
{\it inverse binomial} sums\,\cite{DK04,MKL04}.

The introduction of new functions, such as  {\it multiple polylogarithms}, 
(see Appendix~\ref{zoo}) independently in mathematics and physics\,%
\cite{Broadhurst:1998,Goncharov,Borwein:1999,RV00,GR00},\footnote{
Hyperlogarithms have been considered by 
Kummer, Poincar\'e, and Lappo-Danilevsky; see \cite{Hyper_Historic}. 
The interrelation between hyperlogarithms and multiple polylogarithms has been 
discussed in \cite{Goncharov-Hyperlogarithms}.  
} allows us to derive a set of universal algorithms 
for the simplification and construction of the analytical coefficients of the 
Laurent expansion of a large class of hypergeometric functions. (For details,
see Refs.\ \cite{DK01,MKL06,nested1,nested2,ls1,ls2,hyper:expansion,KWY07}.)
Recently, similar problems have also drawn the attention of 
mathematicians\,\cite{hyper:expansion,SeriesNew}.  However, the general 
solution of this problem remains unknown. 

The multiple series representation has
further applications in the framework of Feynman diagram calculations. 
In particular, the Smirnov-Tausk approach\,\cite{Smirnov,Tausk} 
(see also Ref.\ \cite{ExpMath}) was very productive for constructing the 
analytical coefficients of the $\ep$-expansion (finite part mainly)
of Feynman diagrams depending on one or two massless (ratio of massive) 
kinematic variables. 
Presently, there are several computer realizations of this 
approach\,\cite{Anastasiou}.
In the framework of this technique, the Feynman parameter 
representation\,\cite{Bogolyubov} for a diagram is rewritten in terms of 
multiple Mellin-Barnes (contour integral) representations,
resulting in expressions for which a Laurent expansion about $\ep=0$
may be constructed explicitly, using gamma functions and their derivatives.
The results may be summed analytically or numerically, typically leading
to the same sums as in the construction of the $\ep$-expansion of 
hypergeometric functions: the {\it $($generalized$)$ harmonic sums} and 
{\it $($inverse$)$ binomial} sums. {\it Inverse binomial sums} 
typically arise from massive loops; see Refs.\ \cite{BD,illustration}.
Another source of multiple sums in Feynman diagrams comes from the 
Frobenius series solution of a differential equation \cite{Kotikov}.
Other classes of sums have been considered as well~\footnote{{\it Finite 
harmonic sums} are another class, on which more details may be found in
Ref.\ \cite{finite}. However, there presently does not exist an appropriate 
generalization of {\it multiple (inverse) binomial sums} to finite 
harmonic sums. Some recent attempts in this direction have been discussed  
in Refs.\ \cite{nested2,finite:binomial}.}.

Analytical results are possible when these sums can be evaluated explicitly. 
For the analysis of {\it $($generalized$)$ harmonic sums}, the 
{\it nested sums approach}\,\cite{nested1,nested2} permits the reduction of any
type of {\it $($generalized$)$ harmonic sum} to a set of basis sums.
The analytical evaluation of these basis sums is an independent problem.
(See, for example, Ref.\ \cite{DK04}.)  
The {\it Generating function approach}\,\cite{wilf} is a universal method
for analytically evaluating arbitrary sums, which was successfully applied 
(see section (2.3) in Ref.\ \cite{DK04}) for an analysis of multiple 
(inverse) binomial sums\,\cite{JKV03,DK04}.  The generating function approach 
allows us to convert arbitrary sums to a system of differential equations. 
The question of the expressibility of the solution to this differential 
equation in terms of known (special) functions is not addressed by this approach.
In particular, the partial results of Refs.\ \cite{FKV99,JKV03,DK04,MKL06} 
were restricted by attempts to express the results of the calculation in terms 
of only classical or Nielsen polylogarithms\,\cite{Lewin,Nielsen}.
It is presently unknown what type of sums (beyond generalized harmonic sums) 
are expressible in terms of known special functions.\footnote{
There is not universal agreement on what it means to express a
solution in terms of known special functions.  One reasonable answer has 
been presented by Kitaev in the Introduction to Ref.\ \cite{Kitaev}. 
where he quotes R. Askye's Forward to the book {\it Symmetries and Separation
of Variables} by W. Miller, Jr.,\cite{Miller} which says
``One term which has not been defined so far is `special function'. 
My definition is simple, but not time invariant.  A function is a 
special function if it occurs often enough so that it gets a name''.
Kitaev adds, ``... most of the people who apply them \ldots understand, under 
the notion of special functions, a set of 
functions which can be found in one of the well-known reference books\ldots.''
To this, we may add ``functions which can be found in 
one of the well-known computer algebra systems.''
}

The aim of this paper is to prove the following theorems:

\noindent{\bf Theorem A} \\
{\it 
The multiple $($inverse$)$ binomial sums 
\begin{equation}
\sum_{j=1}^\infty \frac{1}{\binom{2j}{j}^k}\frac{z^j}{j^c}
S_{a_1}(j-1) \cdots S_{a_p}(j-1) \; ,
\label{mbinsum}
\end{equation}
where $k = \pm 1$, $S_a(j) = \sum_{k=1}^j \frac{1}{k^a}$ is a harmonic
series, and $c$ is any integer,
are expressible in terms of Remiddi-Vermaseren functions
with 
\begin{itemize}
\item[] {\rm (1)} for $k=1:$
{\rm (i)}  $c \geq 2$ rational  coefficients; 
{\rm (ii)} $c \leq 1$ ratios of polynomials;

\item[] {\rm (2)} for $k=-1:$
{\rm (i)}  $c \geq 1$ rational  coefficients; 
{\rm (ii)} $c \leq 0$ ratios of polynomials.
\end{itemize}
}

\noindent{\bf Theorem B} \\
{\it 
The all order $\ep$-expansion of the generalized hypergeometric 
functions} \cite{bateman}
\begin{subequations}
\begin{eqnarray}
\label{Theorem2}
& {}_pF_{p-1}&\left(\vec{A}\!+\!\vec{a}\ep; \vec{B}\!+\!\vec{b}\ep, 
\tfrac{1}{2}\!+\! I_1; z\right), 
\label{F:01}
\\ 
& {}_pF_{p-1}&\left(\vec{A}\!+\!\vec{a}\ep, \tfrac{1}{2} 
\!+\! I_2; \vec{B}\!+\!\vec{b}\ep; z\right), 
\label{F:10}
\end{eqnarray}
\end{subequations}
{\it
where $\vec{A}$, $\vec{B}$ are lists of integers and $I_1$, $I_2$ are integers,
are expressible in terms of the harmonic 
polylogarithms with coefficients that are ratios of polynomials.
}

The paper is organised as follows. 
In section \ref{sums}, we will prove {\bf Theorem A}.  In
section \ref{hypergeometric}, the results of {\bf Theorem A} will be 
applied to hypergeometric functions to prove {\bf Theorem B}. 
Section \ref{general_sums} is devoted to a discussion of an algorithm for 
the reduction and analytical evaluation of generalized multiple (inverse) 
binomial sums.  Appendix~\ref{zoo} contains some basic information about 
relevant special functions. 


\section{Analytical evaluation of a basis of multiple (inverse) binomial sums 
of arbitrary weight and depth}
\label{sums}

The main purpose of this section is to prove {\bf Theorem A}.  In the first
subsection, we will consider differential equations satisfied by multiple
(inverse) binomial sums, and use the analytical properties of such sums to
derive two useful lemmas.  In the second subsection, we prove auxiliary
propositions for the separate cases of multiple binomial sums and inverse
binomial sums, and use them to complete the proof of {\bf Theorem A}.

\subsection{Some analytical properties of multiple (inverse) binomial sums 
of arbitrary weight and depth}

Let us define the \emph{multiple sums}
\begin{eqnarray}
\Sigma^{(k)}_{a_1,\cdots,a_p; \; b_1,\cdots,b_q;c}(u)
&\equiv&
\sum_{j=1}^\infty \frac{1}{\binom{2j}{j}^k}\frac{u^j}{j^c}
S_{a_1}(j-1) \cdots S_{a_p}(j-1) S_{b_1}(2j-1) \cdots S_{b_q}(2j-1) \; ,
\nonumber \\
\label{binsum}
\end{eqnarray}
where $S_a(j) = \sum_{k=1}^j \frac{1}{k^a}$ is a harmonic series and 
$c$ is any integer.
For particular values of $k$, the sums (\ref{binsum}) are called 
\begin{eqnarray}
k = 
\left\{ 
\begin{array}{rl}
 0  & \mbox{ {\it generalized harmonic} } \\
 1  & \mbox{ {\it inverse binomial} } \\
-1  & \mbox{ {\it binomial} } 
\end{array} \right\} \mbox{ sums }.
\nonumber 
\end{eqnarray}
The case $\Sigma^{(0)}_{a_1,\cdots,a_p; \; 0,\cdots,0;c}(u)$ is called a 
{\it harmonic} sum.  The number $w=c+a_1+\cdots a_p + b_1 + \cdots b_q$ is 
called the {\bf weight} and  $d=p+q$ is called the {\bf depth}.

The general properties of  multiple sums can be derived from their 
generating functions.  Let us rewrite the multiple sum (\ref{binsum}) 
in the form 
\hbox{$ \Sigma^{(k)}_{\vec{a};\vec{b};c}(u) = \sum_{j=1}^{\infty} u^j 
\eta^{(k)}_{\vec{a};\vec{b};c}(j) \; , $}
where $\vec{a}\equiv \left( a_1,\ldots,a_p\right)$ and $\vec{b}\equiv 
\left(b_1,\ldots,b_q\right)$ denote the collective lists of indices
and  $\eta^{(k)}_{\vec{a};\vec{b};c}(j)$ is the coefficient of $u^j$.
In order to find the differential equation for generating functions of 
{\it multiple sums} it is necessary to find a recurrence relation for the 
coefficients $\eta^{(k)}_{\vec{a};\vec{b};c}(j)$ with respect to the summation
index $j$.  Using the explicit form of $\eta^{(k)}_{\vec{a};\vec{b};c}(j)$,
the recurrence relation for the coefficients can be written in 
the form\footnote{
We would like to point out that Eq.~(\ref{rec:relation}) is valid for an 
arbitrary integer $k$ and $c-k \geq 0$.  In the case $c-k < 0$, the proper 
term will be generated in the r.h.s.\ of the equation.
}
\begin{equation}
\bigl[ 2 (2j+1) \bigr]^k (j+1)^{c-k} \eta^{(k)}_{\vec{a};\vec{b};c}(j+1) 
= j^c \eta^{(k)}_{\vec{a};\vec{b};c}(j) + r^{(k)}_{\vec{a};\vec{b}}(j) \; ,
\label{rec:relation}
\end{equation}
where the ``remainder'' $r^{(k)}_{\vec{a};\vec{b}}(j)$ is given by
\begin{eqnarray}
{
\binom{2j}{j}^k} \; r^{(k)}_{\vec{a};\vec{b}}(j) & = & 
\prod\limits_{r=1}^p \left[ S_{a_r}(j-1) \!+\! j^{-a_r}\right]
\times \prod\limits_{l=1}^q 
\left[ S_{b_l}(2j-1) \!+\! (2j)^{-b_l} \!+\! (2j\!+\!1)^{-b_l} \right]
\nonumber \\ && \hspace{2mm}
- \prod\limits_{r=1}^p \prod\limits_{l=1}^q S_{a_r}(j-1) S_{b_1}(2j-1) \; .
\label{r:AB:1}
\end{eqnarray}

Multiplying both sides of Eq.~(\ref{rec:relation}) by $u^j$,
summing from $j=1$ to $\infty$, and using the fact that any extra power
of $j$ corresponds to the derivative $u({d}/{d}u)$ leads to 
the following differential equations for the generating functions
$\Sigma^{(k)}_{\vec{a};\vec{b};c}(u)$ (see Ref.\ \cite{MKL04}):
\begin{subequations}
\begin{eqnarray}
\label{generating}
&& \hspace{-7mm}
\left[ \left(\tfrac{4}{u} \!-\! 1 \right) u \tfrac{{d}}{{d} u} 
\!-\! \tfrac{2}{u} \right]
\left( u \tfrac{{d}}{{d}u} \right)^{c-1} 
\Sigma^{(1)}_{\vec{a};\vec{b};c}(u)  
= \delta_{p,0} + R^{(1)}_{\vec{a};\vec{b}}(u) \; ,
\label{diff:I}
\\ && \hspace{-7mm}
\left(\tfrac{1}{u} \!-\! 1 \right) \left( u \tfrac{{d}}{{d} u} \right)^c  
\Sigma^{(0)}_{\vec{a};\vec{b};c}(u) =
\delta_{p,0} \!+\! R^{(0)}_{\vec{a};\vec{b}}(u) \; ,
\label{diff:II}
\\ && \hspace{-7mm}
\left[ 
\left(\tfrac{1}{u} \!-\! 4 \right) u \tfrac{{d}}{{d} u}
\!-\! 2 
\right]
\left( u \tfrac{{d}}{{d}u} \right)^{c} 
\Sigma^{(-1)}_{\vec{a};\vec{b};c}(u)  
= 2 \delta_{p,0} + 2 \left( 2 u \tfrac{{d}}{{d} u} + 1\right) R^{(-1)}_{\vec{a};\vec{b}}(u) \; ,
\label{diff:III}
\end{eqnarray}
\end{subequations}
where $R^{(k)}_{\vec{a};\vec{b}}(u)\equiv\sum_{j=1}^{\infty} u^j 
r^{(k)}_{\vec{a};\vec{b}}(j)$ and $\delta_{a,b}$ is the Kronecker 
$\delta$-function.  The boundary conditions for any of these sums and 
their derivatives are 
\begin{equation}
\left(u \frac{d}{du} \right)^j \Sigma_{\vec{a};\vec{b};c}(0) = 0 \;,
\quad j=0,1,2, \cdots
\label{boundary}
\end{equation}

From the analysis in Refs.\ \cite{JKV03,DK04,MKL06,MKL04}, 
we have deduced that the set of equations for the {\it generating functions}
has a simpler form in terms of a new variable. 
For {\it multiple inverse binomial sums}, this variable is defined by
\begin{equation}
y = \frac{\sqrt{u-4}-\sqrt{u}}{\sqrt{u-4}+\sqrt{u}} \;, \quad
u = - \frac{(1-y)^2}{y} \;, \quad  
\label{y}
\end{equation}
and for {\it multiple binomial sums}, it is defined by
\begin{equation}
\chi = \frac{1-\sqrt{1-4u}}{1+\sqrt{1-4u}}, \quad 
u = \frac{\chi}{(1+\chi)^2}\;.
\label{chi}
\end{equation}

Let us consider the differential equation for {\it multiple inverse binomial 
sums} in terms of these new variables.  The notation
$\Sigma^{(k)}_{\vec{a};\vec{b};c}(y)[(\chi)]$ will be used for a sum defined 
by Eq.~(\ref{binsum}), where the variable $u$ is rewritten in terms of 
variable $y[\chi]$ defined by Eq.~(\ref{y}) [(\ref{chi})]:
\begin{eqnarray}
\Sigma^{(1)}_{\vec{a};\vec{b};c}(y) & \equiv &   
\Sigma^{(1)}_{\vec{a};\vec{b};c}\left( u(y) \right) 
\equiv \left.  \Sigma^{(1)}_{\vec{a};\vec{b};c}(u) \right|_{u=u(y)} \;,
\nonumber \\ 
\Sigma^{(-1)}_{\vec{a};\vec{b};c}(\chi) & \equiv &   
\Sigma^{(-1)}_{\vec{a};\vec{b};c}\left( u(\chi) \right) 
\equiv \left.  \Sigma^{(-1)}_{\vec{a};\vec{b};c}(u) \right|_{u=u(\chi)} \;.
\end{eqnarray}

In terms of the variable $y$, equation (\ref{diff:I}) may be split into 
sum of two equations
\begin{subequations}
\label{SIGMA+1}
\begin{eqnarray}
\left( \!-\! \frac{1\!-\!y}{1\!+\!y} y \frac{d}{d y} \right)^{c-1} 
\Sigma^{(1)}_{\vec{a};\vec{b};c}(y) \!&=&\! \frac{1\!-\!y}{1\!+\!y} 
\sigma^{(1)}_{\vec{a};\vec{b}}(y) \; ,
\label{SIGMA+1:a}
\\ 
y \frac{d}{d y} \sigma^{(1)}_{\vec{a};\vec{b}}(y) \!&=&\!\delta_{p,0} 
\!+\! R^{(1)}_{\vec{a};\vec{b}}(y) \; 
\label{SIGMA+1:b}
\end{eqnarray}
\end{subequations}
with boundary condition 
\[
\Sigma^{(1)}_{\vec{a};\vec{b};c}(1) = 0 \;.
\]
Equation (\ref{SIGMA+1:a}) could be rewritten as 
\begin{equation}
\left( \!-\! \frac{1\!-\!y}{1\!+\!y} y \frac{d}{d y} \right)^{c-j} 
\Sigma^{(1)}_{\vec{a};\vec{b};c}(y)  = \Sigma^{(1)}_{\vec{a};\vec{b};j}(y) \;,
\label{rec:y:1}
\end{equation}
or in equivalent form
\begin{equation}
\left( - \frac{1-y}{1+y} y \frac{d}{d y} \right)^{c-j-1} 
\Sigma^{(1)}_{\vec{a};\vec{b};c}(y)  
= \int_1^y dy \left( \frac{2}{1-y} - \frac{1}{y} \right) 
\Sigma^{(1)}_{\vec{a};\vec{b};j}(y) \;.
\label{lemma:A}
\end{equation}
From this representation we immediately obtain the following lemma: \\
\noindent
{\bf Lemma A} (see Ref. \cite{MKL04}) \\
{\it 
If for some integer $j$, the series $\Sigma^{(1)}_{\vec{a};\vec{b};j}(u)$
is expressible in terms of Remiddi-Vermaseren functions {\rm (\ref{HC})} 
with rational coefficients,  then the sums 
$\Sigma^{(1)}_{\vec{a};\vec{b};j+i}(u)$ for positive integers $i$ can also be 
expressed in terms of functions of this type with rational coefficients.
} \\

In a similar manner, let us rewrite the differential equations for the 
generating function of the {\it multiple binomial sums} as
\begin{subequations}
\label{binsum:diff}
\begin{eqnarray}
&& \hspace{-6mm}
\left( \frac{1\!+\!\chi}{1\!-\!\chi} \chi \frac{{d}}{{d} \chi}  \right)^c  
\Sigma^{(-1)}_{\vec{a};\vec{b};c}(\chi)
= \frac{1\!+\! \chi}{1\!-\! \chi} \sigma^{(-1)}_{\vec{a};\vec{b}}(\chi) \; , 
\label{binsumdifI}
\\ && \hspace{-6mm}
\frac{1}{2} (1\!+\!\chi)^2 \frac{{d}}{{d} \chi} 
\sigma^{(-1)}_{\vec{a};\vec{b}}(\chi) \!=\!  \delta_{p,0} \!+\! 
\left(2 \frac{1\!+\! \chi}{1\!-\! \chi} \chi \frac{{d}}{{d} \chi} \!+\! 
1\right) R^{(-1)}_{\vec{a};\vec{b}}(\chi) \; .
\label{binsumdifII}
\end{eqnarray}
\end{subequations}
The first equation may be rewritten as 
\begin{equation}
\left( \frac{1\!+\! \chi}{1\!-\! \chi} \chi \frac{{d}}{{d} \chi} \right)^{c-j} 
\Sigma^{(-1)}_{\vec{a};\vec{b};c}(\chi) 
= \Sigma^{(-1)}_{\vec{a};\vec{b};j}(\chi) \;,
\end{equation}
or in an equivalent form
\begin{equation}
\left( \frac{1\!+\! \chi}{1\!-\! \chi} \chi \frac{{d}}{{d} \chi} \right)^{c-j-1}
\Sigma^{(-1)}_{\vec{a};\vec{b};c}(\chi)  
= \int_0^\chi d \chi \left( \frac{1}{\chi} - \frac{2}{1+\chi} \right) 
\Sigma^{(-1)}_{\vec{a};\vec{b};j}(\chi) \;.
\label{lemma:B}
\end{equation}
In this case, the boundary condition (\ref{boundary}) is unchanged,
and we can make a statement similar to the previous one: \\

\noindent{\bf Lemma B} (see Ref.\ \cite{MKL04}) \\
{\it 
If for some integer $j$, the series $\Sigma^{(-1)}_{\vec{a};\vec{b};j}(u)$
is expressible in terms of {\it harmonic polylogarithms} {\rm (\ref{HC})}
with rational coefficients, then
the sums $\Sigma^{(-1)}_{\vec{a};\vec{b};j+i}(u)$ for positive integers $i$  
can also be expressed in terms of {\it harmonic polylogarithms}
with rational coefficients.
} \\

\subsection{Analytical evaluation of multiple (inverse) binomial sums of 
arbitrary\\ weight and depth}

Let us now consider the special case of sums (\ref{binsum}) including only 
products of harmonic sums, and show that they are expressible in terms of
Remiddi-Vermaseren functions\footnote{
These sums are related to the multiple sums
\[
\sum_{n_1>n_2>\cdots n_p=1}^\infty \frac{1}{\binom{2n_1}{n_1}}
\ \frac{u^{n_1}}{n_1^{c} n_2^{b_1} \cdots n_p^{b_p}} \;.
\]
} with argument $(j-1)$; see Eq.~(\ref{mbinsum}).
In agreement with Ref.\ \cite{DK04}, we will denote such a sum as
$\Sigma^{(k)}_{a_1,\cdots, a_p; \;-;m}(u)$.  In this case, the non-homogeneous 
term $r^{(k)}_{\vec{a};-}(j)$ of differential equation 
(\ref{generating}) is again expressible in terms of sums of the same type, 
$\Sigma^{(k)}_{b_1,\cdots, b_p; \;-;m}(u)$, but with smaller {\bf depth}:
\begin{equation}
{
\binom{2j}{j}^k} \; r^{(k)}_{\vec{a};-}(j) =
\prod\limits_{r=1}^p \left[ S_{a_r}(j-1) \!+\! j^{-a_r}\right]
\!-\!  \prod\limits_{r=1}^p S_{a_r}(j-1) \; , \quad k = \pm 1 \;.
\label{r:AB:2}
\end{equation}

We shall start with the case of {\it inverse binomial sums}, 
$k=1$:
$$
\sum_{j=1}^\infty \frac{1}{\binom{2j}{j}}\frac{u^j}{j^c} S_{a_1}(j-1) 
\cdots S_{a_p}(j-1) 
\;.
$$
In order to prove {\bf Theorem A} for inverse binomial sums, 
we will prove an auxiliary proposition:\\[1em]
\noindent
{\bf Proposition I} \\
{\it For $c=1$, the inverse binomial sums are expressible in terms of harmonic
polylogarithms with rational coefficients $c_{r,\vec{s}}$ times a
factor $(1-y)/(1+y)$: }
\begin{equation}
\left.  \Sigma^{(1)}_{a_1,\cdots, a_p; \;-;1}(u) \right|_{u=u(y)} = 
\frac{1-y}{1+y} \sum_{r,\vec{s}} c_{r,\vec{s}} \ln^r y\ \Li{\binom{\vec{\sigma}}{\vec{s}}}{y} \;,
\label{proposition:1}
\end{equation}
{\it where
$r+s_1+\cdots+s_k=1+a_1+\cdots+a_p$ $(${\bf\emph{weight}} of l.h.s.\ = 
{\bf\emph{weight}} of r.h.s.$)$.}

Substituting expression (\ref{proposition:1}) in the r.h.s.\ of 
Eq.~(\ref{lemma:A}), setting $j=1$, and making trivial splitting of the 
denominator, we get the following result:\\[1em]
\noindent{\bf Corollary A:} \\ {\it
For $c \geq 2$, the inverse binomial sums are expressible in terms of harmonic 
polylogarithms with rational coefficients $d_{r,\vec{s}}$: }
\begin{equation}
\left.  \Sigma^{(1)}_{a_1,\cdots, a_p; \;-;c}(u) \right|_{u=u(y)} = 
\sum_{r,\vec{s}} d_{r,\vec{s}} \ln^r y\ \Li{\binom{\vec{\sigma}}{\vec{s}}}{y}\;,
\quad c \geq 2
\end{equation}
{\it where
$r+s_1+\cdots+s_k=c+a_1+\cdots+a_p$ $(${\bf\emph{weight}} of l.h.s.\ =
{\bf\emph{weight}} of r.h.s.$)$.}\\[1em]

\noindent
{\bf Proof:}\\
Let us consider {\it inverse binomial sums} of {\bf depth 0}:
\[
\Sigma^{(1)}_{-;-;c}(u)
\equiv
\sum_{j=1}^\infty \frac{1}{\binom{2j}{j}}\frac{u^j}{j^c} \;.
\]
It was shown in Ref.\ \cite{KV00} that for any $c \geq 2$ this sum is 
expressible in terms of generalized log-sine functions\,\cite{Lewin} which 
could be rewritten\,\cite{DK01,lsjk} in terms of Nielsen 
polylogarithms.\,\cite{Nielsen} Here we will present an iterated solution for 
the case of interest.  The system (\ref{SIGMA+1}) has the form
\begin{subequations}
\begin{eqnarray}
\left( \!-\! \frac{1\!-\!y}{1\!+\!y} y \frac{d}{d y} \right)^{c-1} 
\Sigma^{(1)}_{-;-;c}(y) \!&=&\! \frac{1\!-\!y}{1\!+\!y} 
\sigma^{(1)}_{-;-}(y) \; ,
\\  
y \frac{d}{d y} \sigma^{(1)}_{-;-}(y) \!&=&\! 1\; .
\end{eqnarray}
\end{subequations}
For $c=1$, we immediately get the relation
\begin{equation}
\Sigma^{(1)}_{-;,-;1}(y) \!=\! \frac{1\!-\!y}{1\!+\!y} \ln y \;,
\label{Sigma:1}
\end{equation}
which coincides  with {\bf Proposition I} and 
can be readily transformed into the form of Eq.~(\ref{rec:y:1}):
\begin{equation}
\left( \!-\! \frac{1\!-\!y}{1\!+\!y} y \frac{d}{d y} \right)^{c-2} \Sigma^{(1)}_{-;-;c}(y)  
= - \frac{1}{2} \ln^2 y \;.
\end{equation}
The iterated solution of this differential equation for an arbitrary integer 
$c \geq 2$ is expressible in terms of Remiddi-Vermaseren functions with 
rational coefficients\footnote{Compare with the results of 
Refs.\ \cite{KV00,DK01,lsjk}.} 
(in accordance with {\bf Corollary A}). 

For sums of {\bf depth 1}, {\it i.e.} 
\[
\Sigma^{(1)}_{a_1;-;c}(u)
\equiv
\sum_{j=1}^\infty \frac{1}{\binom{2j}{j}}\frac{u^j}{j^c} S_{a_1}(j-1) 
\equiv
\sum_{j=1}^\infty \frac{1}{\binom{2j}{j}}\frac{u^j}{j^c} 
\sum_{i=1}^{j-1} \frac{1}{i^{a_1}} \;,
\]
the coefficients of the non-homogeneous part are equal to 
{\it inverse binomial sums} of the zero depth,
\hbox{$ \binom{2j}{j}\; r^{(1)}_{a_1;-}(j) = 1/j^{a_1} \;, $}
and Eqs.~(\ref{SIGMA+1}) take the form
\begin{subequations}
\label{sigma+1:c}
\begin{eqnarray}
&& 
\left( \!-\! \frac{1\!-\!y}{1\!+\!y} y \frac{d}{d y} \right)^{c-1} 
\Sigma^{(1)}_{a_1;-;c}(y) \!=\! \frac{1\!-\!y}{1\!+\!y} \sigma^{(1)}_{a_1;-}(y) \; ,
\label{sigma+1:c1}
\\ && 
y \frac{d}{d y} \sigma^{(1)}_{a_1;-}(y) \!=\! \Sigma^{(1)}_{-;-;a_1}(y)\; .
\label{sigma+1:c2}
\end{eqnarray}
\end{subequations}
For $c=1$ the system of equations (\ref{sigma+1:c}) takes the simplest form
\begin{subequations}
\label{sigma+1:1}
\begin{eqnarray}
&& 
\Sigma^{(1)}_{a_1;-;1}(y) \!=\! \frac{1\!-\!y}{1\!+\!y} 
\sigma^{(1)}_{a_1;-}(y) \; ,
\label{sigma+1:1a}
\\ && 
y \frac{d}{d y} \sigma^{(1)}_{a_1;-}(y) \!=\! \Sigma^{(1)}_{-;-;a_1}(y)\; .
\label{sigma+1:1b}
\end{eqnarray}
\end{subequations}

Let us now consider the case $a_1=1$. Using Eq.~(\ref{Sigma:1}), we derive 
from Eq.~(\ref{sigma+1:1})
\[
\sigma^{(1)}_{1;-}(y) \!=\! \frac{1}{2} \ln^2 y  - 2  \ln y \ln(1+y) 
- 2 \Li{2}{-y}\;, 
\]
{\it i.e.}, result is expressible in terms of harmonic polylogarithms. 
For $a_1 \geq 2$, the r.h.s.\ of the second equation (\ref{sigma+1:1b}) is 
expressible in terms of harmonic polylogarithms with rational coefficients 
(in accordance with previous considerations), so that $\sigma^{(1)}_{a_1;-}(y)$
will also be expressible in terms of harmonic polylogarithms with 
rational coefficients. Substituting these results in the first equation 
(\ref{sigma+1:1a}), we obtain results in accordance with {\bf Proposition I}. 
For $c \geq 2$, the desired result follows from {\bf Lemma A}.

We may complete the proof by mathematical induction. Let us assume that 
{\bf Proposition I} is valid for {\it multiple inverse binomial sums} 
of {\bf depth \emph k}: 
\begin{eqnarray}
\label{+1:k:1}
\Sigma^{(1)}_{a_1, \cdots, a_k;-;1}(u) &\equiv& \left.
\sum_{j=1}^\infty \frac{1}{\binom{2j}{j}}\frac{u^j}{j} S_{a_1}(j-1) 
	\cdots S_{a_k}(j-1)\right|_{u=u(y)}
\nonumber\\
&=&
\frac{1-y}{1+y} \sum_{r,\vec{s}} c_{r,\vec{s}} \ln^r y\ \Li{
\binom{\vec{\sigma}}{\vec{s}}}{y} 
\;,
\end{eqnarray}
where $\Li{\binom{\vec{\sigma}}{\vec{s}}}{z}$
is a coloured polylogarithm of a square root of unity, 
$\vec{s}=s_1,\cdots,s_k$, and $r+s_1+\cdots s_p=c+a_1+\cdots+a_k$.
Then for $c \geq 2$,  {\bf Corollary A} also holds for 
{\it multiple inverse binomial sums} of {\bf depth \emph k}:
\begin{eqnarray}
\Sigma^{(1)}_{a_1, \cdots, a_k;-;c}(u)
\equiv
\left.  \sum_{j=1}^\infty \frac{1}{\binom{2j}{j}}\frac{u^j}{j^c} 
S_{a_1}(j-1) \cdots S_{a_k}(j-1)\right|_{u=u(y)} = 
\sum_{r,\vec{s}} \tilde{c}_{r,\vec{s}} \ln^r y\ 
\Li{\binom{\vec{\sigma}}{\vec{s}}}{y} \;,
\nonumber \\
\label{+1:k:2}
\end{eqnarray}

For the sum of {\bf depth \emph k+1}, the coefficients of the non-homogeneous 
part may be expressed as linear combinations of sums of {\bf depth \emph j}, 
$j=0, \cdots, k$, with integer coefficients and all possible symmetric 
distributions of the original indices between terms of the new sums:
\begin{subequations}
\label{SIGMA+k+1}
\begin{eqnarray}
&& \hspace{-7mm}
\left( \!-\! \frac{1\!-\!y}{1\!+\!y} y \frac{d}{d y} \right)^{c-1} 
\Sigma^{(1)}_{a_1,\cdots , a_{k+1};-;c}(y) \!=\! \frac{1\!-\!y}{1\!+\!y} 
\sigma^{(1)}_{a_1,\cdots , a_{k+1};-}(y) \; ,
\label{SIGMA+k+1:a}
\\ && \hspace{-7mm}
y \frac{d}{d y} \sigma^{(1)}_{a_1,\cdots , a_{k+1};-}(y) \!=\!
\sum_{j=1}^\infty \frac{u^j}{\binom{2j}{j}} 
\sum_{p=0}^{k} \sum_{(i_1,\cdots,i_{k+1})}
\frac{1}{p!(k\!+\!1\!-\!p)!}\frac{S_{i_1}(j\!-\!1) \cdots 
S_{i_p}(j\!-\!1)}{j^{i_{p+1}+\cdots i_{k+1}}}\;,
\label{SIGMA+k+1:b}
\end{eqnarray}
\end{subequations}
where the sum over indices $(i_1,\cdots i_{k+1})$ is to be taken over all
permutations of the list $(a_1, \cdots, a_{k+1})$.
If $i_{p+1}+\cdots i_{k+1} \geq 2$, the r.h.s.\ of Eq.\ (\ref{SIGMA+k+1:b}) 
is expressible in terms of harmonic polylogarithms of {\bf weight \emph k} 
with rational coefficients; see Eq.~(\ref{+1:k:2}).
As the result of integrating this equation, 
$\sigma^{(1)}_{a_1,\cdots , a_{k+1};-}(y)$ also will be expressible in terms 
of harmonic polylogarithms of {\bf weight \emph k+1} with rational coefficients.

If $i_{p+1}+\cdots i_{k+1} = 1$, the r.h.s.\ of Eq.\ (\ref{SIGMA+k+1:b}) 
is expressible in terms of harmonic polylogarithms of {\bf weight \emph k} 
with a common factor $(1-y)/(1+y)$; see Eq.~(\ref{+1:k:1}).
The result of integrating this equation again will be expressible in terms 
of harmonic polylogarithms of {\bf weight \emph k+1} with 
rational coefficients:
\[
\sigma^{(1)}_{a_1,\cdots , a_{k+1};-}(y) = \int_1^y dt 
\left( \frac{1}{t} - \frac{2}{1+t} \right) \sum_{r,\vec{s}} c_{r,\vec{s}} 
\ln^r t\ \Li{\binom{\vec{\sigma}}{\vec{s}}}{t}\;.
\]
For $c=1$, direct substitution of the previous results into
(\ref{SIGMA+k+1:a}) will show that  {\bf Proposition I}
is valid at {\bf weight \emph k+1}. In this way, the {\bf Proposition I} 
is proven for all weights.  Then for $c \geq 2$, {\bf Corollary A} 
is also true for {\it multiple inverse binomial sums} of 
{\bf depth \emph k+1}. \\

Applying the differential operator $u\frac{d}{du} \equiv 
- \frac{1-y}{1+y} y \frac{d}{dy} $ repeatedly $l$ times to the 
sum $\Sigma^{(1)}_{a_1,\cdots,a_p; \;-;c}(u)$, we can derive results for a
similar sum with $c \leq 1$.\footnote{Some particular cases of sums of this 
type were considered also in Ref.\ \cite{Borwein}.} Thus, {\bf Theorem A} is 
proven for {\it multiple inverse binomial sums}.\footnote{All multiple inverse 
binomial sums up to {\bf weight 4} 
were calculated in ref.\ \cite{DK04}; see Table I in Appendix C.}

Let us now consider the {\it multiple binomial sums}\footnote{
These sums are related to the multiple sums
\[
\sum_{n_1>n_2>\cdots n_p=1}^\infty {\binom{2n_1}{n_1}}
\frac{u^{n_1}}{n_1^{c} n_2^{b_1} \cdots n_p^{b_p}} \;.
\]
}, $(k=-1)$, $\Sigma^{(-1)}_{a_1,\cdots,a_p; \;-;c}(u)$
\[
\sum_{j=1}^\infty {\binom{2j}{j}} \frac{u^j}{j^c} S_{a_1}(j-1) 
\cdots S_{a_p}(j-1) \;.
\]

In order to prove {\bf Theorem A} for binomial sums, we will first prove the 
following auxiliary proposition:\\[1em]
\noindent
{\bf Proposition II} \\ \emph{
For $c=0$, the binomial sums are expressible in terms of harmonic 
polylogarithms and have the following structure:}
\begin{equation}
\left.
\Sigma^{(-1)}_{a_1,\cdots, a_p; \;-;0}(u)\right|_{u=u(\chi)}
= \sum_{r,\vec{s}} \Biggl[ \frac{1}{1-\chi} c_{r,\vec{s}} + d_{r,\vec{s}} 
\Biggr]
\ln^r \chi\ \Li{\binom{\vec{\sigma}}{\vec{s}}}{\chi} \;,
\label{proposition:2}
\end{equation}
\emph{where
$r+s_1+\cdots+s_k=1+a_1+\cdots+a_p$ $(${\bf\emph{weight}} of l.h.s.\ =
{\bf\emph{weight}} of r.h.s.$)$ and $c_{r,\vec{s}}$ and 
$d_{r,\vec{s}}$ are rational numbers.}\\[1em]

Substituting the expression (\ref{proposition:2}) in the 
r.h.s.\ of Eq.~(\ref{lemma:B}) and setting $j=0$, we get \\[1em]
\noindent
{\bf Corollary B} \\ \emph{
For $c \geq 1$, the binomial sums are expressible in terms of harmonic 
polylogarithms 
with rational coefficients $\tilde{d}_{r,\vec{s}}$: 
\begin{equation}
\left.  \Sigma^{(-1)}_{a_1,\cdots, a_p; \;-;c}(u)\right|_{u=u(\chi)}
= \sum_{r,\vec{s}} \tilde{d}_{r,\vec{s}} \ln^r 
\chi \ \Li{\binom{\vec{\sigma}}{\vec{s}}}{\chi} \;, \quad c \geq 1 \;, 
\end{equation}
where
$r+s_1+\cdots+s_k=c+a_1+\cdots+a_p$ $(${\bf\emph{weight}} of l.h.s.\ is equal 
to {\bf\emph{weight}} of r.h.s.$)$.}\\[1em]

We start again from the {\it multiple binomial sums} of {\bf depth 0},
\[
\Sigma^{(-1)}_{-;-;c}(u) \equiv
\sum_{j=1}^\infty \binom{2j}{j}\frac{u^j}{j^c} \;.
\]
In this case, Eqs.~(\ref{binsum:diff}) have the form
\begin{subequations}
\label{sigma-1:0}
\begin{eqnarray}
&& 
\left( \frac{1\!+\!\chi}{1\!-\!\chi} \chi \frac{{d}}{{d} \chi}  \right)^c  
\Sigma^{(-1)}_{-;-;c}(\chi)
= \frac{1\!+\! \chi}{1\!-\! \chi} \sigma^{(-1)}_{-;-}(\chi) \; , 
\nonumber \\  && 
\frac{1}{2} (1\!+\!\chi)^2 \frac{{d}}{{d} \chi} \sigma^{(-1)}_{-;-}(\chi) 
\!=\! 1 \;,
\end{eqnarray}
\end{subequations}
where the factor $\frac{1\!+\! \chi}{1\!-\! \chi}$ may be written as 
\begin{equation}
\frac{1\!+\! \chi}{1\!-\! \chi} = \frac{2}{1-\chi} - 1 \;.
\label{factor}
\end{equation}
For $c=0$, we obtain
\begin{equation}
\Sigma^{(-1)}_{-;-;0}(\chi) = 2 \left[ \frac{1}{1-\chi} - 1\right] \;,
\label{sigma:1}
\end{equation}
which coincides with {\bf Proposition II}.
Substituting this result into r.h.s.\ of Eq.~(\ref{lemma:B}) we find
\begin{eqnarray}
\left( \frac{1\!+\!\chi}{1\!-\!\chi} \chi \frac{d}{d \chi} \right)^{c-1} 
\Sigma^{(-1)}_{-;-;c}(\chi)  =  2 \ln (1+\chi) \;.
\label{sigma:0}
\end{eqnarray}
The results of iterated integration, for $c \geq 1$ and boundary condition 
defined by Eq.~(\ref{boundary}), are expressible in terms of {\it generalized 
polylogarithms} (\ref{GP}) with rational coefficients (see Corollary B). 
For the sums of {\bf depth 1},
\[
\Sigma^{(-1)}_{a_1;-;c}(u) \equiv
\sum_{j=1}^\infty \binom{2j}{j}\frac{u^j}{j^c} S_{a_1}(j-1) \equiv
\sum_{j=1}^\infty \binom{2j}{j}\frac{u^j}{j^c} 
\sum_{i=1}^{j-1} \frac{1}{i^{a_1}} \;,
\]
we have 
\begin{subequations}
\label{bin:1}
\begin{eqnarray}
&& 
\left( \frac{1\!+\!\chi}{1\!-\!\chi} \chi \frac{{d}}{{d} \chi}  \right)^c  
\Sigma^{(-1)}_{a_1;-;c}(\chi)
= \frac{1\!+\! \chi}{1\!-\! \chi} \sigma^{(-1)}_{a_1;-}(\chi) \; , 
\label{bin:1:a}
\\  && 
\frac{1}{2} (1\!+\!\chi)^2 \frac{{d}}{{d} \chi} \sigma^{(-1)}_{a_1;-}(\chi) 
\!=\! 
2 \Sigma^{(-1)}_{-;-;a_1-1}(\chi)
+ \Sigma^{(-1)}_{-;-;a_1}(\chi)\;.
\label{bin:1:b}
\end{eqnarray}
\end{subequations}
Integrating by part in Eq.~(\ref{bin:1:b})  and taking into account 
that~\footnote{
This relation follows from the differential relation
$$
u \frac{d}{du} \Sigma^{(k)}_{\vec{a};\vec{b};c}(u)
= 
\Sigma^{(k)}_{\vec{a};\vec{b};c-1}(u) \;.
$$
} 
\[
\frac{d}{d \chi} \Sigma^{(-1)}_{-;-;c}(\chi) = 
\left( \frac{1}{\chi} - \frac{2}{1+\chi}
\right)  \Sigma^{(-1)}_{-;-;c-1}(\chi) \;,
\]
we obtain
\begin{equation}
\sigma^{(-1)}_{a_1;-}(\chi) \!=\! 
-\frac{1-\chi}{1+\chi} \Sigma^{(-1)}_{-;-;a_1}(\chi)
+ \int_0^\chi \frac{dt}{t} \Sigma^{(-1)}_{-;-;a_1-1}(t) \;.
\end{equation}
Using this results in the r.h.s.\ of Eq.~(\ref{bin:1:a}), we have
\begin{equation}
\left( \frac{1\!+\!\chi}{1\!-\!\chi} \chi \frac{{d}}{{d} \chi}  \right)^{c}
\Sigma^{(-1)}_{a_1;-;c}(\chi)
= - \Sigma^{(-1)}_{-;-;a_1}(\chi) + \frac{1+\chi}{1-\chi}
\int_0^\chi \frac{dt}{t} \Sigma^{(-1)}_{-;-;a_1-1}(t) \;.
\label{sigma:c}
\end{equation}
Let us set $c=0$.
It is necessary to consider two cases: (i) $a_1=1$ and (ii) $a_1 \geq 2$.  For 
$a_1=1$, we can use the explicit results (\ref{sigma:1}) and (\ref{sigma:0}) 
to get 
\[
\Sigma^{(-1)}_{1;-;0}(\chi) = 2 \ln (1-\chi) - 2 \ln (1+\chi)
- \frac{4}{1-\chi}  \ln (1-\chi) \;, 
\]
in accordance with {\bf Proposition II}.  For $a_1 \geq 2$ the r.h.s.\ of 
Eq.~(\ref{bin:1:b}) is expressible in terms of harmonic polylogarithms
with rational coefficients, so that Eq.~(\ref{sigma:c}) is also expressible 
in terms of harmonic polylogarithms with rational coefficients in 
accordance with {\bf Proposition II}.  

For $c \geq 1$, the desired result follows from {\bf Lemma B}:
\begin{equation}
\left( \frac{1\!+\!\chi}{1\!-\!\chi} \chi \frac{{d}}{{d} \chi}  \right)^{c-1}
\Sigma^{(-1)}_{a_1;-;c}(\chi) = 
- \Sigma^{(-1)}_{-;-;a_1+1}(\chi) + 
\int_0^\chi \frac{dt_1}{t_1} \int_0^{t_1} \frac{dt_2}{t_2} 
\Sigma^{(-1)}_{-;-;a_1-1}(t_2) \;.
\end{equation}
In particular, for $a_1=1$ we have 
\begin{equation}
\left( \frac{1\!+\!\chi}{1\!-\!\chi} \chi \frac{{d}}{{d} \chi}  \right)^{c-1}
\Sigma^{(-1)}_{1;-;c}(\chi)
= 2 \Li{2}{-\chi} + 2 \ln^2(1+\chi) + 2 \Li{2}{\chi}  \;.
\end{equation}

Let us assume {\bf Proposition II} is valid for 
{\it multiple binomial sums} of {\bf depth \emph k}, 
and prove the proposition for
{\bf depth \emph k+1}. Thus, we assume that
\begin{eqnarray}
\Sigma^{(-1)}_{a_1, \cdots, a_k;-;0}(u) & \equiv & \left.
\sum_{j=1}^\infty \binom{2j}{j} u^j S_{a_1}(j-1) \cdots S_{a_k}(j-1)
\right|_{u=u(\chi)}
\nonumber \\ 
& = & \sum_{p,\vec{s}} \left[ \frac{1}{1-\chi} c_{p,\vec{s}} + d_{p,\vec{s}} 
\right] \ln^p \chi \Li{\binom{\vec{\sigma}}{\vec{s}}}{\chi} \;,
\label{sigma-1:k:0}
\end{eqnarray}
where $\Li{\binom{\vec{\sigma}}{\vec{s}}}{\chi}$
is a coloured polylogarithm of a square root of unity, 
$\vec{s}=(s_1,\cdots,s_k)$, 
and $p+s_1+\cdots s_p=a_1+\cdots+a_k$.
Then for $c \geq 1$, {\bf Corollary B} also holds for 
{\it multiple binomial sums} of {\bf depth \emph k}:
\begin{eqnarray}
\Sigma^{(-1)}_{a_1, \cdots, a_k;-;c}(u) & \equiv & \left.
\sum_{j=1}^\infty \binom{2j}{j} \frac{u^j}{j^c} S_{a_1}(j-1) \cdots 
S_{a_k}(j-1) \right|_{u=u(\chi)} \nonumber \\ 
& = & \sum_{p,\vec{s}} \tilde{c}_{p,\vec{s}} \ln^p \chi 
\Li{\binom{\vec{\sigma}}{\vec{s}}}{\chi} \;,
\label{sigma-1:k:1}
\end{eqnarray}

For a sum of {\bf depth \emph k+1}, the coefficients of the non-homogeneous 
part are expressed as linear combinations of sums of 
{\bf depth \emph j}, $j=0, \cdots, k$, with an integer coefficients and all
possible distributions of the original indices between terms of new sums,
multiplied by a factor $(2j+1)$:
\begin{subequations}
\label{SIGMA-1}
\begin{eqnarray}
&& \left( \frac{1\!+\!\chi}{1\!-\!\chi} \chi \frac{d}{d \chi} \right)^{c} 
\Sigma^{(-1)}_{a_1,\cdots , a_{k+1};-;c}(\chi) \!=\! 
\frac{1\!+\! \chi}{1\!-\! \chi} \sigma^{(-1)}_{a_1,\cdots , a_{k+1};-}(\chi)\; ,
\label{SIGMA-1:a}
\\ && 
\frac{1}{2} (1\!+\!\chi)^2 \frac{{d}}{{d} \chi} \
\sigma^{(-1)}_{a_1,\cdots , a_{k+1};-}(\chi) \!=\!  \sum_{j=1}^\infty (2j+1)
\binom{2j}{j} u^j
\nonumber \\ && \hspace{2cm}
\times \sum_{p=0}^{k} \sum_{(i_1,\cdots,i_{k+1})} \frac{1}{p!(k+1-p)!}
\frac{S_{i_1}(j\!-\!1) \cdots S_{i_p}(j\!-\!1)}{j^{i_{p+1}+\cdots i_{k+1}}} \;,
\label{SIGMA-1:b}
\end{eqnarray}
\end{subequations}
where the sum over indices $(i_1,\cdots, i_{k+1})$ is to be taken over all
permutations of the list $(a_1, \cdots, a_{k+1})$.

Let us denote the sub-list of length $p$ as $\vec{I} =(i_1, \cdots, i_p)$ and 
define the sum of the remaining indices as $J=i_{p+1}+\cdots +i_{k+1}$, 
so that the second equation (\ref{SIGMA-1:b}) can be written as 
\[
\frac{1}{2} (1\!+\!\chi)^2 \frac{{d}}{{d} \chi} 
\sigma^{(-1)}_{a_1,\cdots , a_{k+1};-}(\chi) 
\!=\!  \sum_{\vec{I},J} \left[
2 \Sigma^{(-1)}_{\vec{I};-;J}(\chi) + 
\Sigma^{(-1)}_{\vec{I};-;J-1}(\chi) \right] \;.
\]
Integrating by parts, we find
\[
\sigma^{(-1)}_{a_1,\cdots , a_{k+1};-}(\chi) = \sum_{\vec{I},J}
\left[ -\frac{1-\chi}{1+\chi} \Sigma^{(-1)}_{\vec{I};-;J}(\chi)
+ \int_0^\chi \frac{dt}{t} \Sigma^{(-1)}_{\vec{I};-;J-1}(t) \right] \;.
\]
Substituting this result into the r.h.s.\ of Eq.~(\ref{SIGMA-1:a}) we have 
\begin{equation}
\left( \frac{1\!+\!\chi}{1\!-\!\chi} \chi \frac{d}{d \chi} \right)^{c} 
\Sigma^{(-1)}_{a_1,\cdots , a_{k+1};-;c}(\chi) = \sum_{\vec{I},J}
\left[ - \Sigma^{(-1)}_{\vec{I};-;J}(\chi) + \frac{1+\chi}{1-\chi}
\int_0^\chi \frac{dt_1}{t_1} \Sigma^{(-1)}_{\vec{I};-;J-1}(t_1)
\right] \;.
\label{SIGMA-1:k+1}
\end{equation}

Let us set $c=0$ and consider two cases: 
(i) $J=1$ and (ii) $J \geq  2$.
For $J =1$, the first term of the r.h.s.\ of Eq.~(\ref{SIGMA-1:k+1})
is expressible in terms of harmonic polylogarithms with rational coefficients. 
The last term of the r.h.s.\ of Eq.~(\ref{SIGMA-1:k+1}) has the structure of 
Eq.~(\ref{sigma-1:k:0}) so that after integration, it will again be
expressible in terms of harmonic polylogarithms of {\bf weight \emph k+1}. 
For $J \geq 2$, both terms of the r.h.s.\ of Eq.~(\ref{SIGMA-1:k+1}) are 
expressible in terms of harmonic polylogarithms of {\bf weight \emph k+1} 
(see Eq.~(\ref{SIGMA-1:k+1})).  In this way, the {\bf Proposition II} is 
found to be valid at the {\bf weight \emph k+1}. Consequently, 
{\bf Proposition II} is proven for all weights. Therefore, for $c \geq 1$, 
{\bf Corollary B} is also valid for the {\it multiple binomial sums} 
of {\bf weight \emph k+1}.

Applying the differential operator 
$u\frac{d}{du} = \frac{1+\chi}{1-\chi} \chi \frac{d}{d \chi}$ repeatedly 
$l$ times to the sum $\Sigma^{(-1)}_{a_1,\cdots,a_p; \;-;c}(\chi)$, 
we can derive results for similar sums with $c \leq 0$.
Thus, {\bf Theorem A} is proven for {\it multiple binomial sums}.
\footnote{All multiple binomial sums up to {\bf weight 3} 
were calculated in ref.\ \cite{JKV03,DK04}; see the proper Appendixes.}

\section{All-order $\ep$-expansion of hypergeometric functions 
with one half-integer value of the parameters via multiple (inverse) 
binomial sums}
\label{hypergeometric}
In this section, we turn our attention to the proof of {\bf Theorem B}. 
It is well known that any function 
${}_{p}F_{p-1}(\vec{a}+\vec{m};\vec{b}+\vec{k}; z)$ is expressible 
in terms of $p$ other functions of the same type:
\begin{eqnarray}
&& \hspace{-5mm}
R_{p+1}(\vec{a},\vec{b},z) {}_{p}F_{p-1}(\vec{a}+\vec{m};\vec{b}+\vec{k}; z) = 
\sum_{k=1}^{p}R_k(\vec{a},\vec{b},z) {}_{p}F_{p-1}(\vec{a}+\vec{e_k};\vec{b}
+\vec{E_k}; z) \;,
\label{decomposition}
\end{eqnarray}
where $\vec{m},\vec{k}, \vec{e}_k,$ and $\vec{E}_k$ are lists of integers and
$R_k$ are polynomials in parameters $\vec{a},\vec{b}$, and $z$.
Systematic methods for solving this problem were elaborated in 
Refs.\ \cite{zeilberger,takayama}.  For generalized hypergeometric functions 
of {\bf Theorem B}, let us choose as basis 
functions arbitrary $p$-functions from the following set: 
\begin{itemize}
\item
for Eq.~(\ref{F:01}) there are $p^2$ functions of the proper type: 
\[
_{p}F_{p-1}\left(\begin{array}{c|}
\tfrac{3}{2}, \{ 1+a_i\ep\}^{p-L-1}, \; \{ 2+d_i\ep\}^L  \\
\{ 1+e_i\ep \}^{p-Q-1}, \{ 2+c_i\ep \}^Q
\end{array} ~z \right) \;,
\]
\item
for Eq.~(\ref{F:10}) there are $p^2-1$ functions of the proper type: 
\[
_{p}F_{p-1}\left(\begin{array}{c|}
\{ 1+a_i\ep\}^{p-L}, \; \{ 2+d_i\ep\}^L  \\
\tfrac{3}{2}, \{ 1+e_i\ep \}^{p-Q-2}, \{ 2+c_i\ep \}^Q 
\end{array} ~z \right) \;.
\]
\end{itemize}

In the framework of the approach developed in 
Refs.\ \cite{KV00,DK01,JKV03,DK04,MKL04}, 
the study of the $\ep$-expansion of basis hypergeometric functions has been 
reduced to the study of multiple {\it $($inverse$)$ binomial} sums.
It is easy to get the following representations:
\begin{subequations}
\label{hyper0}
\begin{eqnarray}
\hspace{-6mm}{}_{p}F_{p-1}\left(\begin{array}{c|}
\{ 1+a_i\ep\}^K, \; \{ 2+d_i\ep\}^L  \\
\tfrac{3}{2}, \{ 1+e_i\ep \}^R, \{ 2+c_i\ep \}^Q 
\end{array} ~z \right)
&=& \frac{1}{2z} \frac{ \Pi_{s=1}^Q (1+c_s\ep)} {\Pi_{i=1}^{L} (1+d_i\ep)}
\sum_{j=1}^\infty \frac{1}{\binom{2j}{j} }  \frac{(4z)^j}{j^{K-R-1}} \Delta \;, 
\label{inversebinomial1}
\\  
\hspace{-6mm}{}_{p}F_{p-1}\left(\begin{array}{c|}
\tfrac{3}{2}, \{ 1\!+\!a_i\ep\}^K, \; \{ 2\!+\!d_i\ep\}^L  \\
\{ 1 \!+\!e_i\ep \}^R, \{ 2\!+\!c_i\ep \}^Q
\end{array} ~  z \right)
&=& \frac{2}{z} \frac{ \Pi_{s=1}^Q (1\!+\!c_s\ep) }{ \Pi_{i=1}^{L}   
(1\!+\!d_i\ep) } \sum_{j=1}^\infty \binom{2j}{j} 
\frac{\left( \frac{z}{4} \right)^j}{j^{K-R-1}} \Delta \; ,
\label{binomial1}
\end{eqnarray}
\end{subequations}
where the superscripts $K,L,R,Q$ show the lengths of the parameter lists, 
\begin{equation}
\Delta = 
\exp \left[ \sum_{k=1}^{\infty} \frac{(-\ep)^k}{k}  
\left( w_k j^{-k} + S_k(n-1) t_k \right) \right] =  1 
- \ep \left( \frac{w_1}{j} +  t_1 S_1(n-1) \right) + {\cal O}(\ep^2)\; , 
\label{expansion1}
\end{equation}
$S_a(n) = \sum_{j=1}^n 1/j^a$ is a harmonic sum,
and the constants are defined as 
\begin{eqnarray*}
&& \hspace*{-12mm}
A_k \equiv \sum a_i^k, \quad 
C_k \equiv \sum c_i^k, \quad 
D_k \equiv \sum d_i^k, \quad 
E_k \equiv \sum e_i^k, \quad
\label{acde}
\\
&& 
t_k \equiv C_k + E_k - A_k - D_k , \quad
w_k \equiv C_k - D_k \; , 
\end{eqnarray*}
where the summations extend over all possible values of the parameters
in Eqs.~(\ref{hyper0}).  In this way, the $\ep$-expansions of the basis 
functions (\ref{hyper0}) are expressible in terms of 
{\it multiple $($inverse$)$ binomial sums} studied in 
Sect.~\ref{sums}. But all these are are expressible in terms of 
harmonic polylogarithms.  Thus, {\bf Theorem B} is proven.

\section{Generalized multiple (inverse) binomial sums via derivatives of \\
 generalized hypergeometric functions }
\label{general_sums}

In physical applications, in particular, within Smirnov-Tausk approach,
more general sums, in addition to the ones defined in Eq.~(\ref{mbinsum}), 
may be generated:
\[
\sum_{j=1}^\infty \left[ \frac{(j+c_1)!(j+c_2)!}{(2j+c_3)!} \right]^k
\frac{u^j}{(nj+c_4)^c} S_{a_1}(m_1j+b_1) \cdots S_{a_k}(m_kj+b_k) \;,
\]
where $\{a_i\},\{b_j\},\{c_k\}, \{m_k\},n $ are integers and $k=\pm 1$. 
The procedure of 
finding the proper differential equation (see Refs.\ \cite{DK04,wilf} for a 
detailed discussion) can be applied to analytically evaluate 
any of these new sums.  Another approach is based on extension of the 
algorithm of nested sums\,\cite{nested1,nested2}
for the study of the algebraic relations between these sums.
However, there is a third approach arising from the possibility of reducing an 
arbitrary generalized hypergeometric function to a set of basis functions 
with the help of the Zeilberger-Takayama algorithm described by
Eq.~(\ref{decomposition}). 

To be more specific, let us divide both sides of Eq.~(\ref{decomposition})
by $R_{p+1}(a_i,b_j,z)$ and construct the $\ep$-expansion for the 
hypergeometric functions described in {\bf Theorem B}.  The r.h.s.\ of this 
relation is expressible analytically in terms of harmonic polylogarithms with 
polynomial coefficients.  The l.h.s.\ can be used as a generating function for 
generalized multiple (inverse) binomial sums.  Using a standard form for 
the Taylor expansion of the Gamma function,\footnote{
The relation between harmonic sums $S_a(j)$
and derivatives of the function $\psi(z)=\frac{d}{{d}z}\ln\Gamma(z)$ is
\[
\psi^{(k-1)}(j) = (-1)^k (k-1)! \left[\zeta_k - S_k(j-1) \right],
  \qquad  k>1.
\]
} 
\[
\frac{(m+a\ep)_j}{(m)_j} =
\exp\left\{ -\sum_{k=1}^{\infty} \frac{(-a\ep)^k}{k}
\left[ S_k(m \!+\! j \!-\! 1) \!-\! S_k(m \!-\! 1) \right] \right\} \; ,
\]
where 
$(\alpha)_j\equiv\Gamma(\alpha+j)/\Gamma(\alpha)$ is the Pochhammer symbol, 
we obtain
\begin{eqnarray}
&& \hspace*{-7mm}
{}_{P\!+\!1}F_P \left(\begin{array}{c|} 
\{ m_l\!+\!a_l\ep \}^{L}, \{ p_i\!+\! \tfrac{1}{2} \}^{P\!+\!1\!-\!L} \\ 
\{ n_k\!+\!b_k\ep \}^K,  \{ q_j\!+\! \tfrac{1}{2} \}^{P\!-\!K}  
\end{array} ~z \right) 
= \sum_{j=0}^\infty
\frac{z^j}{j!} \frac{\Pi_{l=1}^L(m_l+a_l\ep)_j}{\Pi_{l=1}^K(n_k+b_k\ep)_j} 
\frac{\Pi_{i=1}^{P+1-L}\left(p_i+\frac{1}{2}\right)_j}{\Pi_{s=1}^{P-K}\left(q_s+\frac{1}{2}\right)_j}
\nonumber \\ && \hspace*{-3mm}
= 
\sum_{j=0}^\infty \frac{z^j}{j!} \frac{1}{4^{j(K\!-\!L\!+\!1)}} 
\frac{\Pi_{l=1}^L (m_l)_j}{\Pi_{k=1}^K (n_k)_j}
\prod_{i=1}^{P\!+\!1\!-\!L} \frac{(2 p_i\!+\!1)_{2j}}{(p_i\!+\!1)_{j}}
\prod_{s=1}^{P\!-\!K} \frac{(l_s\!+\!1)_j}{(2 l_s\!+\!1)_{2j}}\;
\Delta \; , 
\label{hyper}
\end{eqnarray}
where the $m_l,n_k,p_i,q_j$ are integers and 
\begin{eqnarray}
&& 
\Delta  =  
\exp \Biggl[ \sum_{k=1}^{\infty} \frac{(-\ep)^k}{k}  
\Biggl( 
\sum_{\omega=1}^K b_\omega^k \left[ S_k(n_\omega\!+\!j\!-\!1) \!-\! S_k(n_\omega\!-\!1) \right]
\nonumber \\ &&  \hspace{25mm}
- \sum_{i=1}^L \left[ a_i^k S_k(m_i\!+\!j\!-\!1) \!-\! a_i^k S_k(m_i\!-\!1) \right] 
\Biggr) 
\Biggr] \; . 
\nonumber 
\end{eqnarray}
Setting $K=L=P$ in Eq.~(\ref{hyper}), we get
generating functions for generalized multiple binomial sums: the derivatives
\begin{subequations}
\label{gbs}
\begin{equation}
\left.
\prod_{l,k} 
\left( \frac{\partial}{\partial a_l} \right)^{r_l}
\left( \frac{\partial}{\partial b_k} \right)^{s_k}
{}_{P\!+\!1}F_P \left(\begin{array}{c|} 
\{ \!a_l \}^P,  p \!+\! \tfrac{1}{2} \\ 
\{ \!b_k \}^P
\end{array} ~z \right) 
\right|_{a_l=m_l; b_k=n_k}
\end{equation}
lead to terms in the epsilon expansion of the form
\begin{equation}
\sum_{j=0}^\infty 
\frac{(2 p\!+\!1)_{2j}}{(p\!+\!1)_{j}}
\frac{1}{j!} \frac{z^j}{4^{j}} \frac{\Pi_{l=1}^P (m_l)_j}{\Pi_{k=1}^P (n_k)_j} 
\prod_{M=1} S_{a_M}(I_M\!+\!j) \; ,
\end{equation}
\end{subequations}
where the $I_M$ are integers from the lists $\{m_l\}^L$ and $\{n_k\}^K$.
For $L=P+1$ and $K=P-1$ we get generating functions for 
generalized multiple inverse binomial sums: 
\begin{eqnarray}
&& 
\left.
\prod_{l,k} 
\left( \frac{\partial}{\partial a_l} \right)^{r_l}
\left( \frac{\partial}{\partial b_k} \right)^{s_k}
{}_{P\!+\!1}F_P \left(\begin{array}{c|} 
\{ a_l \}^{P+1} \\ 
\{ b_k \}^{P-1}, q \!+\! \tfrac{1}{2}
\end{array} ~z \right) 
\right|_{a_l=m_l; b_k=n_k}
\Rightarrow
\nonumber \\ && 
\qquad \qquad
\sum_{j=0}^\infty 
\frac{(q\!+\!1)_{j}}{(2 q\!+\!1)_{2j}}
\frac{(4z)^j}{j!} \frac{\Pi_{l=1}^{P+1} (m_l)_j}{\Pi_{k=1}^{P-1} (n_k)_j} 
\prod_{M=1}  S_{a_M}(I_M\!+\!j) \; .
\label{gibs}
\end{eqnarray}
For $K=P$ and $L=P+1$ we get generating functions for 
generalized multiple harmonic sums: 
\begin{eqnarray}
&& 
\left.
\prod_{l,k} 
\left( \frac{\partial}{\partial a_l} \right)^{r_l}
\left( \frac{\partial}{\partial b_k} \right)^{s_k}
{}_{P\!+\!1}F_P \left(\begin{array}{c|} 
\{ \!a_l \}^{P+1} \\ 
\{ \!b_k \}^{P}
\end{array} ~z \right) 
\right|_{a_l=m_l; b_k=n_k}
\Rightarrow
\nonumber \\ && 
\qquad \qquad
\sum_{j=0}^\infty 
\frac{1}{j!} \frac{\Pi_{l=1}^{P+1} (m_l)_j}{\Pi_{k=1}^P (n_k)_j} 
\prod_{M=1} S_{a_M}(I_M\!+\!j) \; .
\label{ghs}
\end{eqnarray}

Instead of one hypergeometric function, we could consider 
a linear combination of the functions of the same type. 
Such a combination is also reducible and expressible in terms 
of our basis functions. Combining the proper set of hypergeometric functions, 
we could expect that any individual sums,\footnote{Using the results of 
the all-order $\ep$-expansion for Gauss hypergeometric 
functions\,\cite{MKL06,KWY07} we could consider a series of 
type (\ref{binsum}).} of the type  described by r.h.s.\ of 
Eqs.~(\ref{gbs}) -- (\ref{ghs}) are expressible in terms of generalized 
(harmonic) polylogarithms with polynomial coefficients.\footnote{
In particular, all sums presented in Ref.\ \cite{riemann} are reducible in 
terms of our basis sums or sums studied in Ref.\ \cite{DK04}.
Indeed, taking into account that 
$$
(2n+1) \left(2n \atop n\right) = 
\frac{n+1}{2} 
\left(2n+2 \atop n+1\right) 
$$
and shifting the index of summation we have
\begin{eqnarray}
&& 
\sum_{n=1}^\infty 
\frac{1}{\left(2n \atop n \right)}\frac{z^n}{(2n\!+\!1)}X_{\vec{a}}(n)Y_{\vec{b}}(2n\!+\!1) 
= 
\frac{2}{z} 
\sum_{j=1}^\infty 
\frac{1}{\left(2j \atop j \right)}\frac{z^j}{j}X_{\vec{a}}(j\!-\!1)Y_{\vec{b}}(2j\!-\!1) 
- 
X_{\vec{a}}(0)Y_{\vec{b}}(1) \;, 
\label{1}
\end{eqnarray}
where $X_{\vec{a}}(n) = \Pi_{k=1}^r S_{a_k}(n)$
and $Y_{\vec{a}}(2n+1) = \Pi_{k=1}^r S_{a_k}(2n+1)$, 
are products of harmonic sums, with
the vector $\vec{a}$ having $r$ components.
As a consequence, $X_{\vec{a}}(0) = 0$ and  $Y_{\vec{b}}(1) = 1$.
In this way, any sums described by Eq.~(\ref{1}) may be reduced to sums
of type (\ref{binsum}), and for $Y_{\vec{b}}(j)=1$ they are reduced to 
the sums studied in the present paper. 
Another possible generalization of the sums considered here is 
\begin{eqnarray}
&& 
\sum_{n=1}^\infty 
\frac{1}{\left(2n \atop n \right)}\frac{z^n}{(2n\!+\!1)}X_{\vec{a}}(n+1)Y_{\vec{b}}(2n\!+\!1) 
= 
\frac{2}{z} 
\sum_{j=1}^\infty 
\frac{1}{\left(2j \atop j \right)}\frac{z^j}{j}X_{\vec{a}}(j)Y_{\vec{b}}(2j\!-\!1) 
- 
X_{\vec{a}}(1)Y_{\vec{b}}(1) \;. 
\label{2}
\end{eqnarray}
Due to the depth reduction relation, 
$$
X_{\vec{a}}(j) = X_{\vec{a}}(j-1) + 
\sum_{p=0}^{r} \sum_{(i_1,\cdots,i_{k+1})}
\frac{1}{p!(r-p)!}
\frac{S_{i_1}(j\!-\!1) \cdots S_{i_p}(j\!-\!1)}{j^{i_{p+1}+\cdots i_{k+1}}} \;,
$$
sums of type (\ref{2}) are also expressible in terms sums of type (\ref{1}).
}

These arguments suggest a criterion for what type of 
generalized multiple (inverse) binomial sum are expressible 
in terms of harmonic polylogarithms with coefficients that are 
ratios of polynomials.  This is just the beginning of a general analysis, 
but the corresponding analysis for harmonic sums is already known to be 
valid.\,\cite{nested1}  Unfortunately, existing computer 
algebra algorithms\,\cite{a=b} do not allow us to identify the multiple 
series with derivatives of hypergeometric functions or their combinations.
It is still matter of personal experience, 
but this approach looks very promising and is worthy of further analysis. 

\section{Discussion and Conclusions}

We have constructed an iterative solution for {\it multiple $($inverse$)$ 
binomial sums} defined by Eq.~(\ref{mbinsum}). It was shown that by the 
appropriate change of variables, defined by Eqs.~(\ref{y}) and (\ref{chi}), 
the multiple (inverse) binomial sums are converted into harmonic 
polylogarithms (see {\bf Theorem A}).
Symbolically, this may be expressed as
\begin{subequations}
\begin{eqnarray}
\left.
\sum_{j=1}^\infty \frac{1}{\binom{2j}{j}}\frac{u^j}{j} S_{a_1}(j-1) 
\cdots S_{a_k}(j-1) \right|_{u=u(y)}\hspace{-5mm} & = & 
\frac{1-y}{1+y} \sum_{p,\vec{s}} c_{p,\vec{s}} \ln^p y
\ \Li{\binom{\vec{\sigma}}{\vec{s}}}{y} \,,
\\ 
\left.
\sum_{j=1}^\infty \frac{1}{\binom{2j}{j}}\frac{u^j}{j^c} S_{a_1}(j-1) \cdots 
S_{a_k}(j-1)\right|_{u=u(y)}\hspace{-5mm} & = & 
\sum_{p,\vec{s}} \tilde{c}_{p,\vec{s}} \ln^p y\ 
\Li{\binom{\vec{\sigma}}{\vec{s}}}{y} \,, \quad c \geq 2
\end{eqnarray}
\label{map1}
\end{subequations}
and 
\begin{subequations}
\begin{eqnarray}
\hspace{-3mm}\left.
\sum_{j=1}^\infty \binom{2j}{j} u^j S_{a_1}(j-1) \cdots S_{a_k}(j-1)
\right|_{u=u(\chi)}\hspace{-5mm} & = & \sum_{p,\vec{s}} 
\left[ 
\frac{c_{p,\vec{s}}}{1-\chi} 
+ 
d_{p,\vec{s}} 
\right]
\ln^p \chi\ \Li{\binom{\vec{\sigma}}{\vec{s}}}{\chi} \,,
\\ 
\hspace{-3mm}\left.  \sum_{j=1}^\infty \binom{2j}{j} \frac{u^j}{j^c} 
S_{a_1}(j-1) \cdots S_{a_k}(j-1) \right|_{u=u(\chi)}\hspace{-5mm}
& = & \sum_{p,\vec{s}} \tilde{c}_{p,\vec{s}} \ln^p 
\chi\ \Li{\binom{\vec{\sigma}}{\vec{s}}}{\chi} \,, \quad c \geq 1
\end{eqnarray}
\label{map2}
\end{subequations}
where $c$ is a positive integer, 
$ c_{p,\vec{s}}$, $\tilde{c}_{p,\vec{s}}$ and $d_{p,\vec{s}}$ are rational 
coefficients, the {\bf weight} of l.h.s.\ = {\bf weight} of r.h.s., 
$\Li{\binom{\vec{\sigma}}{\vec{s}}}{\chi}$ is 
the coloured multiple polylogarithm of a square root of unity,
\[
S_a(j-1) = \sum_{i=1}^{j-1} \frac{1}{i^a} \;,
\]
is a harmonic series.  The mappings (\ref{map1}), (\ref{map2}) are defined 
in the radius of convergence of the l.h.s.: 
\begin{eqnarray}
\label{convergene}
|u| \leq \left\{
\begin{array}{l}
4, \hspace{1mm}  \; \quad \mbox{inverse binomial} \\
\frac{1}{4} ,
\hspace{1mm} \; \quad \mbox{binomial} 
\end{array}
\right.
\end{eqnarray}

Unfortunately, one of the unsolved problem is the completeness of 
the representation (\ref{map1}), (\ref{map2}). In other words, is it possible 
to express all harmonic polylogarithms in terms of multiple (inverse) binomial 
sums? If not, what kind of sums must be added to get a complete basis? 
Another problem beyond our present considerations is to find 
the algebraic relations among the sums.  

From representation (\ref{map1}), (\ref{map2}), it is evident that 
some (or all, if the basis is complete) of the alternating or 
non-alternating~\footnote{Let us recall that multiple Euler-Zagier sums are
defined as
\begin{equation}
\zeta(s_1,\ldots, s_k; \; \sigma_1,\ldots, \sigma_k) 
= \sum_{n_1>n_2> \ldots >n_k>0}\;\;\; \prod_{j=1}^{k}
\frac{(\sigma_j)^{n_j}}{n_j^{s_j}},
\label{euler}
\end{equation}
where $\sigma_j=\pm 1$ and $s_j>0$. 
$\sigma=1$ is called non-alternating and 
$\sigma=-1$ is alternating sums, correspondingly.
} multiple Euler-Zagier 
sums (or multiple zeta values)\,\cite{Euler-Zagier},
can be written in terms of multiple 
(inverse) binomial sums of special values of arguments. 
Two arguments where such a representation is possible are trivially 
obtained by setting the 
arguments of the harmonic polylogarithms $y, \chi$ to $ \pm 1$:
\begin{eqnarray}
u & = & 4 \;, \quad y =  -1 \;,
\\
u & = & \frac{1}{4} \;, \quad \chi =  1\;.
\end{eqnarray}
Another such point~\footnote{We are thankful to Andrei Davydychev for 
information about the relation between this point and the 
``golden ratio'',\,\cite{Lewin},
$\frac{3-\sqrt{5}}{2} = \left( \frac{1-\sqrt{5}}{2} \right)^2.$
} 
\begin{equation}
u =  -1 \;, \quad y =  \frac{3-\sqrt{5}}{2}  \;
\end{equation}
has been discussed intensively in the context of Ap\'ery-like 
expressions for Riemann zeta functions (see \cite{apery} and References therein). For two other points 
\begin{eqnarray}
u & = & 1 \;, \quad y = \exp \left( i \frac{\pi}{3} \right) \;,
\\
u & = & 2 \;, \quad y =  i \;,
\end{eqnarray}
the relation between 
multiple inverse binomial sums and multiple zeta values was analysed mainly 
by the method of experimental mathematics.\,\cite{pslq} Some of the relations 
are presented in Ref.\ \cite{Borwein:2000} and in the appendix of 
Ref.\ \cite{DK01}.

Let us make a few comments about harmonic polylogarithms of a complex argument.
For the case $0 \leq u \leq 4$, the variable $y$ defined in (\ref{y}) belongs 
to a complex unit circle, $y=\exp (i \theta)$. 
In this case, the coloured polylogarithms of a square root of unity 
can be split into real and imaginary parts as in the case of classical 
polylogarithms.\,\cite{Lewin} 
At the present, there is no commonly accepted notation
for the new functions generated by such splitting. 
In Ref.\ \cite{Borwein:2000}, the {\it multiple Glaishers} and 
{\it multiple Clausen } functions were introduced as the real and 
imaginary parts of generalized polylogarithms of complex unit argument.  
In Ref.\ \cite{DK01,ls1,ls2,lsjk}, the splitting of Nielsen 
polylogarithms was analysed in detail.  In this case, the real and imaginary 
parts are reduced to classical Clausen functions, $\Cl{j}{\theta}$
and generalized log-sine functions $\LS{j}{k}{\theta}$.
Ref.\ \cite{imaginary} attempts to classify new functions on the basis of 
new $\LsLsc{i}{j}{k}{\theta}$-functions. 

In Appendix A of Ref.\ \cite{DK04}, the iterated representation for 
Remiddi-Vermaseren functions of complex unit was constructed. 
It was observed\,\cite{KV00,DK01,DK04,FK99} that the physically interesting 
case, representing single-scale diagrams with with two massive particle cuts, 
corresponds to Remiddi-Vermaseren functions (\ref{color}) with argument equal 
to a primitive ``sixth root of unity'', $y=\exp\left( i \frac{\pi}{3} \right)$. 
This gives an explanation of the proper ``basis of transcendental constants'' 
constructed in Refs.\ \cite{FK99} and \cite{DK01}, and its difference from 
the proper basis of  Broadhurst \cite{Broadhurst:1998}.
Of course, for numerical evaluation of harmonic polylogarithms of complex 
argument, only a series representation is necessary.\cite{numeric}

Using the results of {\bf Theorem A}, we have proved {\bf Theorems B} about 
the all-order $\ep$-expansion of a special class of hypergeometric functions. 
The proof includes two steps: (i) the algebraic reduction of generalized 
hypergeometric functions of the type specified in {\bf Theorems B} to basic 
functions and (ii) the algorithms for calculating the analytical
coefficients of the $\ep$-expansion of basic hypergeometric functions.
The implementation of step (i) -- the reduction algorithm -- is based on 
general considerations performed in Refs.\ \cite{zeilberger,takayama}.
In step (ii), the algorithm is based on series representation of the basis 
hypergeometric functions defined by Eq.~(\ref{hyper}).
The coefficients of the $\ep$-expansion are expressible in terms of 
multiple (inverse) binomial sums analyzed in  {\bf Theorem A}.

Exploring the opportunity to reduce an arbitrary generalized hypergeometric 
function to a set of basis functions with the help of the Zeilberger-Takayama 
algorithm, we have presented in section \ref{general_sums} some arguments
about one possible generalization of (inverse) binomial sums
(see Eq.~(\ref{ghs})) which would be expressible in terms of harmonic 
polylogarithms with coefficients that are ratios of polynomials. 

\acknowledgments 

We are indebted to A.\ Davydychev, A.\ Kotikov, H.~Gangl for interesting 
discussions.  
We would like to thank 
A.~Kotikov, T.~Riemann and O.~Tarasov  for carefully reading the manuscript
and for pointing out some typos in the first version of paper
and A.~Davydychev for checks of some formulae.
This research was supported by NATO Grant PST.CLG.980342 and 
DOE grant DE-FG02-05ER41399.  
Kalmykov is supported in part by BMBF 05 HT6GUA.
M.Yu.K. is thankful to Baylor University for 
support of this research and very grateful to his wife, Laura Dolchini, for 
moral support while working on the paper.\\[1em]

\appendix
\noindent{\large\bf Appendix:}\\[-1em]
\section{Zoo of special functions}\hfill
\label{zoo}

For completeness, we will present the definition of a set of new 
functions, such as {\it multiple polylogarithms}~\footnote{For a review, we 
recommended Ref.\ \cite{mpl}.}
\begin{equation}
\Li{k_1,k_2, \cdots, k_n}{z_1,z_2,\cdots, z_n} = 
\sum_{m_1 > m_2 > \cdots m_n > 0} \frac{z_1^{m_1} z_2^{m_2} 
\cdots z_n^{m_n} }{m_1^{k_1} m_2^{k_2} \cdots m_n^{k_n}} \;.
\label{goncharov}
\end{equation}
Special cases of {\it multiple polylogarithms}\footnote{Our notations corresponds to 
Waldschmitd's paper of Ref.\ \cite{mpl}.} include {\it generalized polylogarithms},
defined by
\begin{equation}
\Li{k_1,k_2, \cdots, k_n}{z} = 
\sum_{m_1 > m_2 > \cdots m_n > 0} \frac{z^{m_1}}{m_1^{k_1} m_2^{k_2} 
\cdots m_n^{k_n}} \;,
\label{GP}
\end{equation}
and {\it coloured polylogarithms of a square root of unity}, defined 
by\footnote{We call $n$ {\bf depth}, and 
$k=k_1+k_2+\cdots+k_n (s=s_1+s_2+\cdots+s_n)$ the {\bf weight}.}
\begin{equation}
\Li{\binom{\vec{\sigma}}{\vec{s}}}{z} \equiv 
\Li{\binom{\sigma_1, \sigma_2, \cdots, \sigma_k}{s_1, s_2, \cdots, s_n}}{z} 
= \sum_{m_1 > m_2 > \cdots m_n > 0} z^{m_1} \frac{\sigma_1^{m_1} \cdots 
\sigma_n^{m_n} }{m_1^{s_1} m_2^{s_2} \cdots m_n^{s_n}} \;,
\label{coloured}
\end{equation}
where  
$\vec{s}=(s_1, \cdots s_n)$ and $\vec{\sigma} = (\sigma_1, \cdots, \sigma_n)$
are multi-indices and $\sigma_k$ is a square root of 
unity, $\sigma_k = \pm 1$.  The extension of {\it coloured polylogarithms of 
square root of unity} (\ref{coloured}) by inclusion of powers of logarithms, 
$\ln^k z$, leads to {\it harmonic polylogarithms} 
or Remiddi-Vermaseren polylogarithms (or functions)\,\cite{RV00}.
These can written in the following form:
\begin{subequations}
\begin{eqnarray}
H_{\vec{A}}(z) & = & \sum_{p,\vec{k }} c_{p,\vec{k}} 
\ln^p z\ \Li{k_1,k_2, \cdots, k_n}{z} \;, 
\label{HG}
\\ 
H_{\vec{B}}(z) & = & \sum_{p,\vec{s}} c_{p,\vec{s}} 
\ln^p z\ \Li{\binom{\vec{\sigma}}{\vec{s}}}{z} \;.
\label{HC}
\end{eqnarray}
\label{RV}
\end{subequations}
where in the first equation, (\ref{HG}), the vector $\vec{A}$ includes only 
components $0$ and $1$, and in the second, (\ref{HC} ), $-1$ components are
included.  The coefficients $c_{\vec{k}}$ and $c_{\vec{s}}$ are rational 
numbers. In Eq.~(\ref{RV}) the {\bf\emph{weight}} of the l.h.s.\  =
the {\bf\emph{weight}} of the r.h.s.

Recall that {\it generalized polylogarithms} (\ref{GP}) can be expressed as 
iterated integrals of the form
\begin{eqnarray}
\Li{k_1, \cdots, k_n}{z} & = &  \int_0^z 
\underbrace{\frac{dt}{t} \circ \frac{dt}{t} \circ \cdots \circ 
\frac{dt}{t}}_{k_1-1 \mbox{ times}} \circ \frac{dt}{1-t} \circ \cdots \circ 
\underbrace{\frac{dt}{t} \circ \frac{dt}{t} \circ \cdots \circ 
\frac{dt}{t}}_{k_n-1 \mbox{ times}} \circ \frac{dt}{1-t} \;,  
\label{iterated}
\end{eqnarray}
where, by definition
\begin{eqnarray}
\int_0^z \underbrace{\frac{dt}{t} \circ \frac{dt}{t} \circ \cdots \circ 
\frac{dt}{t}}_{k_1-1 \mbox{ times}} \circ \frac{dt}{1-t} 
= 
\int_0^z 
\frac{dt_1}{t_1} \int_0^{t_1} \frac{dt_2}{t_2} \cdots 
\int_0^{t_{k-2}} \frac{dt_{k_1-1}}{t_{k_1-1}}
\int_0^{t_{k_1-1}} \frac{dt_{k_1}}{1-t_{k_1}} \;.
\end{eqnarray}
The integral (\ref{iterated}) is an iterated Chen integral \cite{Chen} 
w.r.t.\ the differential forms $\omega_0 = dz/z$ and 
$\omega_1 = \frac{dz}{1-z}$, so that 
\begin{eqnarray}
\Li{k_1, \cdots, k_n}{z} & = & \int_0^z \omega_0^{k_1-1} \omega_1 
\cdots \omega_0^{k_n-1} \omega_1 \;.
\label{chen}
\end{eqnarray}
The coloured polylogarithms (Eq.\ \ref{coloured})) also have
an iterated integral representation w.r.t.\ three differential forms, 
\begin{eqnarray}
\omega_0 & = & \frac{dy}{y}, \quad \sigma=0,
\nonumber \\ 
\omega_\sigma & = & \frac{\sigma dy}{1- \sigma y}, \quad \sigma= \pm 1,
\end{eqnarray}
so that
\begin{equation}
\Li{\binom{\sigma_1, \sigma_2, \cdots, \sigma_k}{s_1, s_2, \cdots, s_k}}{y} 
= \int_0^1 
\omega_0^{s_1-1} \omega_{\sigma_1}
\omega_0^{s_2-1} \omega_{\sigma_1 \sigma_2}
\cdots
\omega_0^{s_k-1} \omega_{\sigma_1 \sigma_2 \cdots \sigma_k}
\;,  \quad \sigma_k^2 = 1\;.
\label{color}
\end{equation}


\end{document}